\documentclass{iau} 
\usepackage{graphicx}

\newcommand{\arcmin}{$^{\prime}$}
\newcommand{\arcsec}{$^{\prime\prime}$}

\newcommand{\kms}{km\,s$^{-1}$}

\newcommand{\fil}{$f_\mathrm{IL}$}
\newcommand{\fnl}{$f_\mathrm{NL}$}
\newcommand{\wil}{$W_\mathrm{IL}$}
\newcommand{\wnl}{$W_\mathrm{NL}$}
\newcommand{\dil}{$\Delta \mu_\mathrm{IL}$}
\newcommand{\dnl}{$\Delta \mu_\mathrm{NL}$}

\newcommand{\Ha}{H$\alpha$}
\newcommand{\ghafas}{\texttt{GH$\alpha$FaS}}

\title[BDSs in Tycho's NE rim] %% give here short title %%
{Balmer-dominated shocks in Tycho's SNR: omnipresence of CRs}

\author[Kne\v{z}evi\'{c}, L\"asker, van de Ven, et al.]   %% give here short author list %%
{Sladjana Kne\v{z}evi\'{c}$^1$, Ronald L\"asker$^2$, Glenn van de Ven$^3$, Joan Font$^4$, John C. Raymond$^5$, Coryn A.~L. Bailer-Jones$^3$, John Beckman$^4$, Giovanni Morlino$^6$, Parviz Ghavamian$^7$, John P. Hughes$^8$
 \and Kevin Heng$^9$}

\affiliation{$^1$ Department of Particle Physics and Astrophysics, Faculty of Physics, The Weizmann Institute of Science, P.O.Box 26, Rehovot 76100, Israel \\email: {\tt sladjana.knezevic@weizmann.ac.il}\\
             $^2$ Finnish Centre for Astronomy with ESO (FINCA), University of Turku, V\"ais\"al\"antie 20, FI-21500 Kaarina, Finland\\
             $^3$ Max Planck Institute for Astronomy, K\"{o}nigstuhl 17, D-69117, Heidelberg, Germany\\
             $^4$ Instituto de Astrof\'{i}sica de Canarias, V\'{i}a L\'{a}ctea, La Laguna, Tenerife, Spain\\
             $^5$ Harvard-Smithsonian Center for Astrophysics, 60 Garden Street, Cambridge, MA 02138, U.S.A.\\
             $^6$ INFN – Gran Sasso Science Institute, viale F. Crispi 7, 67100 L'Aquila, Italy\\
             $^7$ Department of Physics, Astronomy and Geosciences Towson University, Towson, MD 21252, U.S.A.\\
             $^8$ Department of Physics and Astronomy, Rutgers University, 136 Frelinghuysen Road, Piscataway, NJ 08854, U.S.A.\\
             $^9$ University of Bern, Center for Space and Habitability, Sidlerstrasse 5, CH-3012, Bern, Switzerland}

\pubyear{2017}
\volume{331}  %% insert here IAU Symposium No.
\jname{SN 1987A, 30 years later--Cosmic Rays and Nuclei from Supernovae and their aftermaths}
\editors{M. Renaud, A. Marcowith, G. Dubner, A. Ray \& A. Bykov, eds.}
\begin{document}

\maketitle

\begin{abstract}
We present wide-field, spatially and highly resolved spectroscopic observations of Balmer filaments in the northeastern rim of Tycho's supernova 
remnant in order to investigate the signal of cosmic-ray (CR) acceleration. The spectra of Balmer-dominated shocks (BDSs) have characteristic narrow (FWHM $\sim$\,10\,\kms) and broad (FWHM $\sim$\,1000\,\kms) \Ha\ components. CRs affect the H$\alpha$-line parameters: heating the cold neutrals 
in the interstellar medium results in broadening of the narrow H$\alpha$-line width beyond 20\,\kms, but also in reduction of the broad H$\alpha$-line width due to energy being removed from the protons in the post-shock region. For the first time we show that the width of the narrow H$\alpha$ line, much larger than 20\,\kms, is not a resolution or geometric effect nor a spurious result of a neglected intermediate (FWHM $\sim$\,100\,\kms) component resulting from hydrogen atoms undergoing charge exchange with warm protons in the broad-neutral precursor. Moreover, we show that a narrow line width $\gg$\,20\,\kms\ extends across the entire NE rim, implying CR acceleration is ubiquitous, and making it possible to relate its strength to locally varying shock conditions. Finally, we find several locations along the rim, where spectra are significantly better explained (based on Bayesian evidence) by inclusion of the intermediate component, with a width of 180\,\kms\ on average. 
\end{abstract}

\firstsection % if your document starts with a section,
\section{Introduction}

Balmer-dominated shocks (BDSs) are observed around many supernova remnants (SNRs) as bright edge-on sheets of shocked gas. 
BDSs are non-radiative, collisionless and relatively high velocity ($\geqslant$\,200\,\kms) shocks interacting with dilute ($\sim$\,1\,cm$^{-3}$) partially 
ionized medium (\cite{heng10}). Spectroscopically BDSs exhibit bright Balmer lines, in particular narrow (NL) and broad (BL) H$\alpha$ components (\cite{ckr80}). Both components originate downstream of the shock in radiative decays of either pre-shock hydrogen atoms that have been excited in the 
shock or have undergone charge exchange (CE) reactions with the post-shock protons. The latter process produces a so-called broad-neutrals.
In an idealized scenario, the pre-shock gas is not affected (e.g. temperature, ionization state) by the oncoming shock until the latter arrives. 
However, emission from the post-shock gas can influence the physical conditions in the pre-shock region. This phenomenon is termed a precursor. The shocked
gas can be a source of photons, particles and waves that can overtake the shock and heat the pre-shock gas. Mach number, ambient density and 
orientation of the magnetic field define the precursor type. Its existence can affect the lines' widths, intensities and even their shapes.

Balmer filaments are very faint ($\sim$\,10$^{-16}$\,erg\,s$^{-1}$\,cm$^{-2}$\,arcsec$^{-2}$). This requires an instrument with a very good efficiency. Furthermore, they are usually extended ($\sim$\,1\arcmin) and have complex structures, thus we need a large field-of-view (FOV) to cover them in their entirety
and a high spatial resolution is desirable to analyze individual structures of the shock. Finally, to be able to resolve both components, i.e. $\sim$\,10\,\kms\ NL width and $\sim$\,1000\,\kms\ BL width, a high spectral resolution and a large-enough spectral coverage to include all of the BL are needed. At the moment, there is no a single instrument that satisfies all these requirements.  

Tycho's SNR is one of remnants famous for its prominent Balmer filaments. They have been imaged at high spatial resolution with the Hubble Space Telescope (\cite{lee10}). However, spectroscopically only a few bright, small regions in these filaments were observed, and remained spatially unresolved (\cite{ghava00,lee07}). In this paper, we will present spatially resolved observations of the northeastern (NE) filaments that cover and spectrally resolve the narrow component. As an additional step forward, our analysis employs Bayesian inference to reliably quantify confidence intervals and to compare models by means of the evidence for possibly multiple narrow- and intermediate-line (IL width $\sim$\,100\,\kms) components. 

\section{Data and Analysis}

With the FOV of 3.4\arcmin $\times$ 3.4\arcmin\ and the pixel scale of 0.2\arcsec\ of the Galactic H$\alpha$ Fabry-P\'erot Spectrometer (\ghafas) on the William Herschel Telescope (WHT) we covered the entire Tycho's NE rim and spatially resolved its Balmer filaments. The \ghafas\ spectral coverage of 392\,\kms\ and spectral dispersion of 8.1\,\kms\ (the instrument profile is Gaussian, see \cite{blasco10}) precluded simultaneous analysis of the broad component. 
We modified the standard procedure for \ghafas\ data reduction described by \cite{hernandez08} to reduce the $\approx$\,9.6\,h observations. We carried out wavelength and phase calibration for each individual exposure, from which we obtained individual data-subcubes. We did not use the optical derotator so that we could use the largest possible \ghafas\ FOV. Therefore, before co-adding the data-subcubes, we aligned and derotated them. After co-addition of the subcubes, we obtained a final datacube of 48 calibrated constant-wavelength channels. Following the same procedure, we created a background and a flatfield cubes.
We modeled the background flux and flatfield image in individual exposures, and subsequently processed the individual background and flatfield frames in exactly the same manner as the corresponding data frames. 

The seeing of 1\arcsec\ sets the lower size of the binned area that we want to analyze to 19 pixels. Following this requirement, we binned pixels of the NE filaments to get 82 Voronoi spatial-spectral bins (\cite{cc03}) with nearly equal signal-to-noise ($S/N$) of 10. Furthermore, we excluded bins that cover area larger than 400 pixels, so that the estimated $\approx$\,2\% residual background variations do not significantly affect our measurements. 

Our goal is to find a model that we define as $M(\theta) = S(\theta) \times F+B$, where $S$ characterize the shock emission, $F$ flatfield, $B$ background, and its parameters $\theta$ that best describe the data. We model the shock emission with a Gaussian profile to account for a NL, while in our small spectral window, the BL is sufficiently well approximated by a constant component. Presence of a precursor introduces either a wider NL and/or a split in the NL, or an IL (we assume that all lines have a Gaussian profile), depending on a precursor's type. Heating and momentum transfer in a cosmic-ray (CR) precursor result in a NL component wider than the maximal thermal width of 20\,\kms, but also in a split in the NL if we have two shocks inclined with the respect to the line-of-sight (LOS). CE in a broad-neutral (BN) precursor would produce an IL. BN precursor, created by fast BNs that overtake the shock, is expected in shocks with velocities of 2500\,\kms\ (as in the NE rim of Tycho; \cite{ghava01}) that propagate in a partially neutral medium, and formation of a CR precursor is expected 
if the acceleration process in the shock is efficient.

To account for possible NL split or IL presence, we apply several parametrized models to each data set (bin and spectrum): two single-NL models (NL and NLIL), and two double-NL models (NLNL and NLNLIL). To find which of the four models and their parameters best describe the data, we perform Bayesian analysis.
Bayes' theorem defines the \textit{posterior} probability density function (PDF) as a function of model $M$'s parameters $\theta$:
$ P(\theta|D,M) =  P(D|\theta,M) P(\theta|M)/P(D|M)$, where $P(\theta|M)$ is the \textit{prior} -- parameter PDF that we either assume or know before taking data $D$ into account, $P(D|\theta,M)$ is the \textit{likelihood} which is the probability of the data for a given model and parameter set, and $P(D|M)$ is the {\it evidence} for model M (the posterior marginalized over all parameters). Our detector follows a Poisson distribution which we defined to be the likelihood.
We use Markov chain Monte Carlo (MCMC) to draw samples from the posterior and use an ensemble sampling method, see \cite{fmackey13} and \cite{gw10} for details on the sampling algorithm. The evidence integral is often computationally expensive, the reason we approximated it with the \textit{leave-one-out cross validation} (LOO-CV) likelihood (\cite{cbj12}). The base-10 logarithmic evidence ratios are compared and this way we identify the favored model. Although we cannot exclude any of the models, we favor the model with the highest evidence having log evidence ratio relative to the NL model larger than 0.5\,dex (factor $\approx$\,3). If there is no model that is at least 3 times more likely than the NL model, we take the NL model as the favored one. 

\section{Results}

\begin{figure*}[t!]
	%\hspace{-0.3cm}
	%\vspace{-0.2cm}
		\includegraphics[width=0.75\textwidth,height=\textwidth,angle=-90]{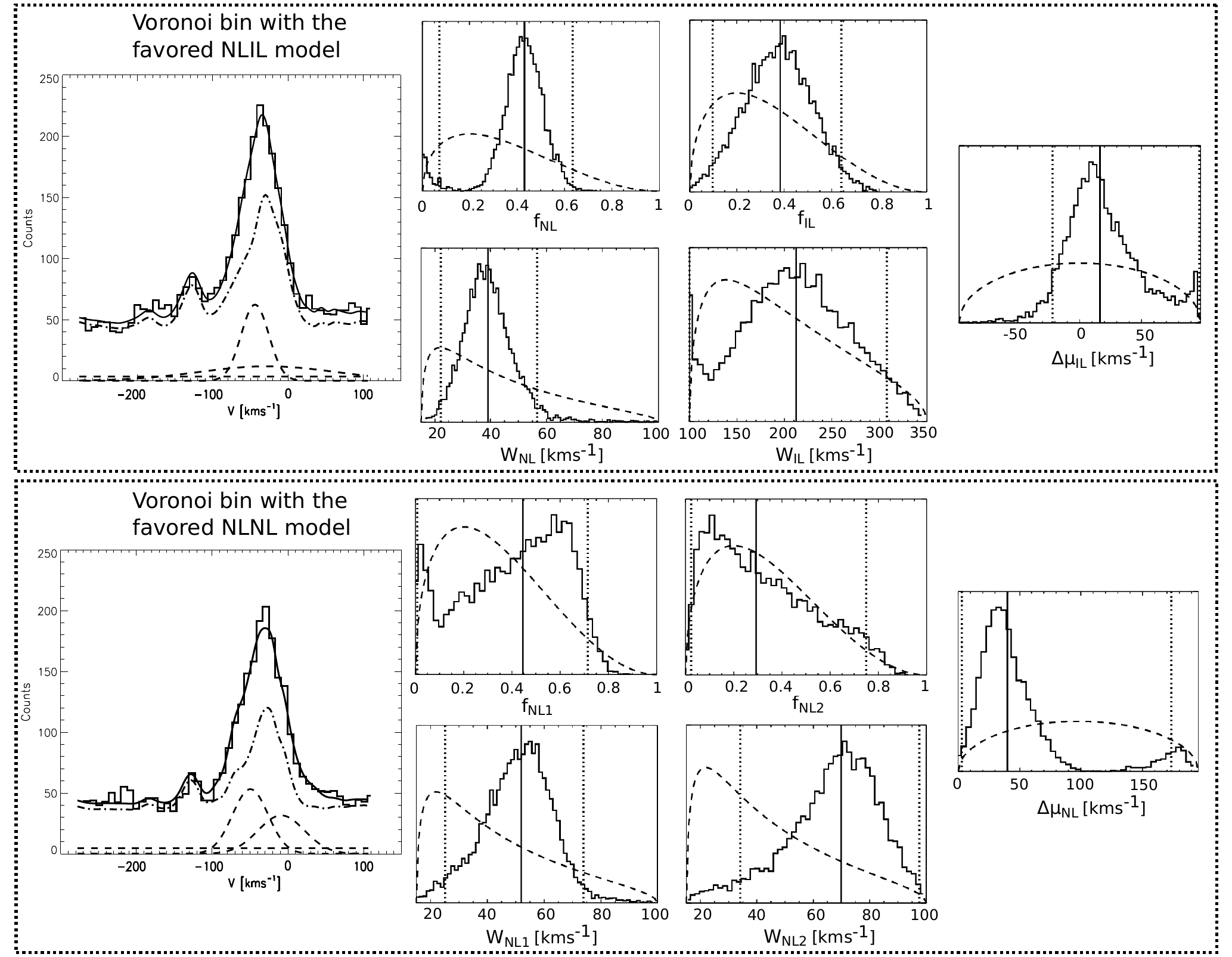}
	\caption{\small Voronoi bins in the NE filament of Tycho's SNR with the NLIL model (\textit{top row}), and the NLNL model (\textit{bottom row}) being the favored one. Left panels show the data as histograms, background models as dashed-dotted lines, dashed lines are the shock emission components, and the median models $M$ are solid lines. The remaining five panels are marginalized posteriors of individual model parameters (solid lines): flux fractions in the lines (\fnl, \fil, $f_\mathrm{NL1}$, $f_\mathrm{NL2}$), intrinsic line widths (\wnl, \wil, $W_\mathrm{NL1}$, $W_\mathrm{NL2}$), IL offset from the NL centroid (\dil), and the separation between the two NLs (\dnl). Overplotted dashed lines are prior distributions. Solid vertical lines are medians of the posterior distributions, and the vertical dotted lines are the 95\% confidence intervals.}
	\label{fig:fig1}
\end{figure*}

\begin{figure*}[t!]
	\centering
		\includegraphics[width=0.65\textwidth,angle=-90]{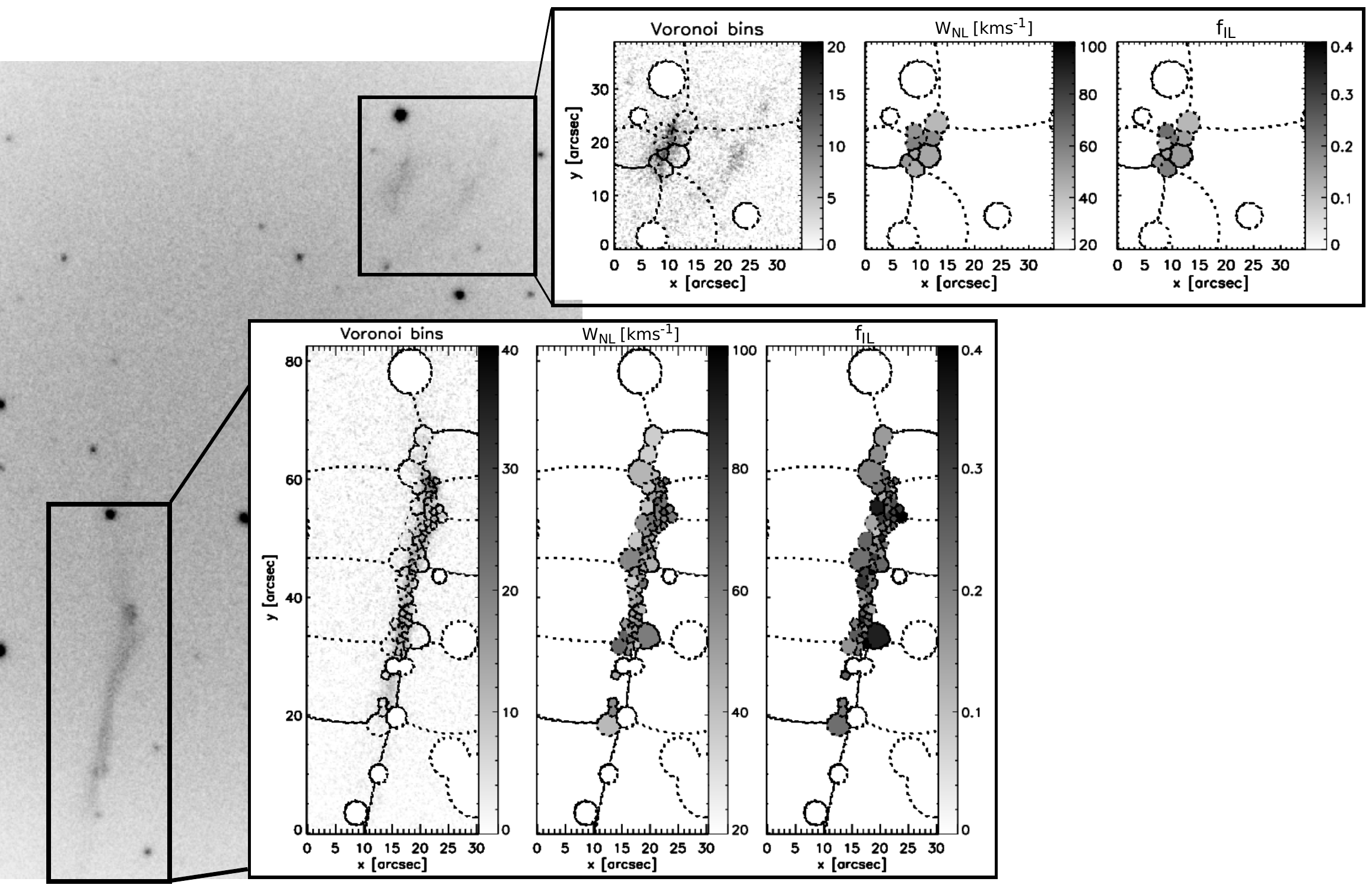}
%\vspace{-0.2cm}
	\caption{\small NE Balmer filaments in Tycho's SNR. The two boxes on the \ghafas\ \Ha\ image show the northern and eastern shock filaments. 
	The panels in each box represent: the bin contours overplotted on the background subtracted \ghafas\ image, spatial variation of the 
	median values of evidence-weighted NL width (in \kms) and IL flux fraction posteriors.}
	\label{fig:fig2}
\end{figure*}

\begin{figure*}[t!]
	\centering
		\includegraphics[width=0.57\textwidth,angle=-90]{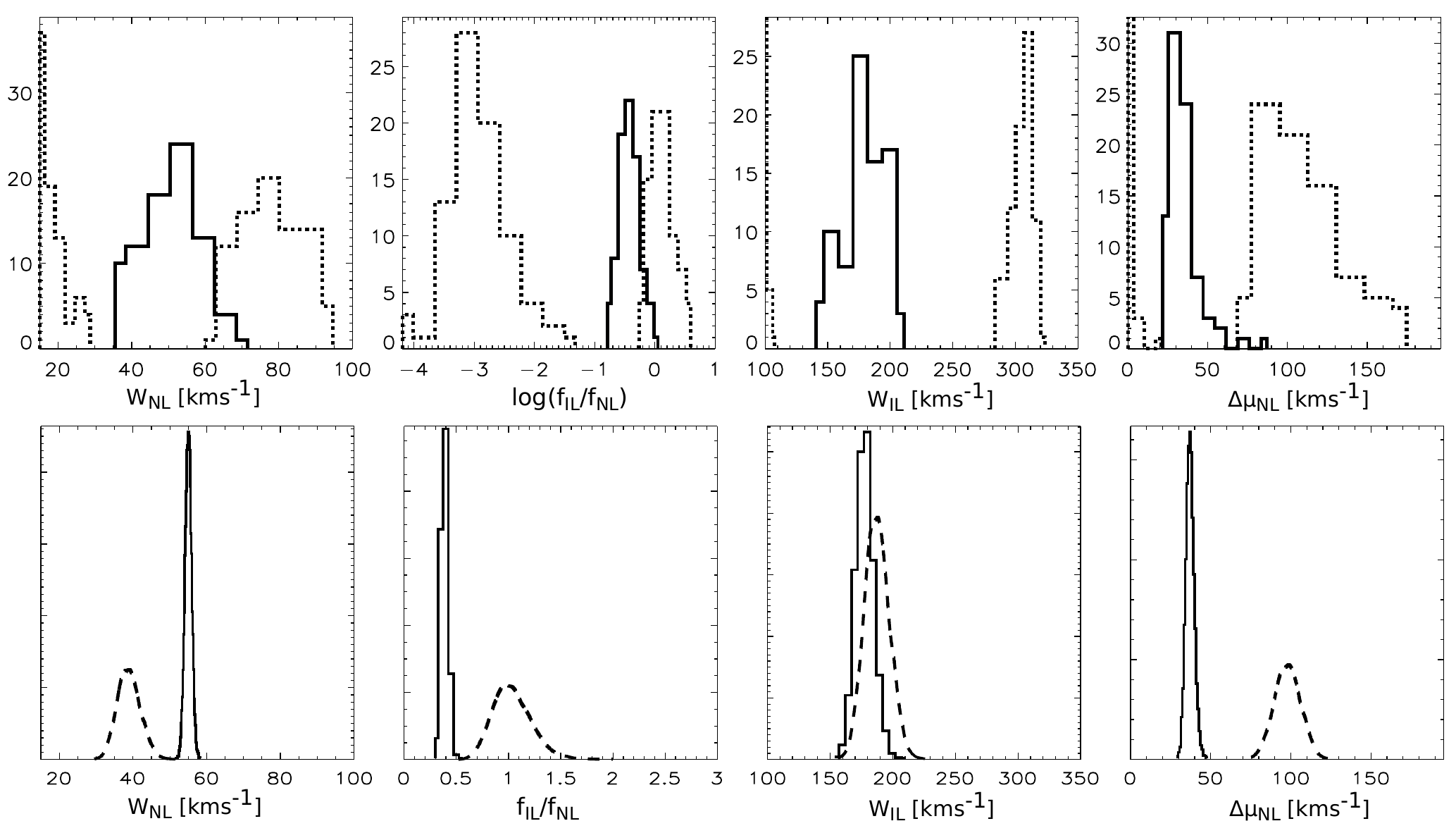}
\vspace{-0.3cm}
	\caption{\small \textit{Top row}: Distribution of median (solid) and highest-density 95\%-confidence interval boundaries (dotted), quantifying the variation of narrow-line width (\wnl, left panel), intermediate-to-narrow line flux fraction (\fil/\fnl, centre-left), IL width (\wil, centre-right) and NL-centroid separation (\dnl, right-most panel) across the filaments. \textit{Bottom}: Posterior of the median across all bins (solid) created using all data to constrain the respective parameter values. Prior of the cross-bin median is overplotted with a dashed line. }
	\label{fig:fig3}
\end{figure*}

We summarize the results in three ways: i) For every Voronoi bin and model we calculated posterior parameter distributions, and give their median and the highest-density 95\% confidence intervals (later on 95\% confidence interval) (Figure~\ref{fig:fig1}); 
ii) For every Voronoi bin we created evidence-weighted 1D-posterior of all models that feature the parameter of interest so that they contribute their marginalized posterior in proportion to their evidence, and calculated their median (Figure~\ref{fig:fig2}) and the 95\% confidence interval (top row in Figure~\ref{fig:fig3});
iii) We consider the parameter's median across bins, sample its posterior (bottom row in Figure~\ref{fig:fig3}), and give this posterior distribution's median and confidence intervals as we did for the individual bins' posteriors. This provides us a measure of the global parameter values when considering the filament as a whole.  
We are particularly interested in the measured NL width, evidence for IL, its strength and width, evidence for a split in the NL, and the separation between the two NLs.

In Figure~\ref{fig:fig1} we present two different bins, one favoring the NLIL model with log evidence ratio larger than 1\,dex compared to the NL and NLNL model, and the other favoring the NLNL model by $>$\,1\,dex (factor 10) over the NL and $>$\,0.5\,dex (factor 3) over the NLIL and NLNLIL models. 
Left panels show the data (histogram) and the median model (solid line), while the remaining panels show priors of chosen parameters (dashed lines), posteriors 
(solid line), the median (solid-vertical line) and the 95\% confidence intervals (dotted-vertical line) of the posteriors. For the NL's full-width at half-maximum (FWHM) $W_\mathrm{NL}$ we used a prior range of [15,\,100]\,\kms\ that reflects the pre-shock temperature of $\approx$\,5000\,K and the maximally predicted $W_\mathrm{NL}$ based on the \cite{mor3i13} shock model that includes the emission from the CR precursor. We limited the $W_\mathrm{IL}$ prior to [100,\,350]\,\kms\ which is expected for the shock velocities around 2000\,\kms\ (\cite{mor2i12}). We do not strongly prefer any parameter values within the model definition limits, which we ensured by using Dirichlet and Beta prior distributions with $\alpha$(= $\beta$) = 1.5 shape parameter for the model parameters (flux fractions in the lines, log-line widths, NL centroid, NL-centroid separation and IL offset from the NL centroid). For the bin with the favored NLIL model we find $W_\mathrm{NL} \approx$\,40\,\kms, and an IL component with $W_\mathrm{IL} \approx$\,210\,\kms\ and $\approx$\,40\% of the total flux. The intrinsic widths of the two NLs in the bin with the favored NLNL model are also much larger than 20\,\kms, nearly 52\,\kms\ and 70\,\kms, separated by 40\,\kms.

In Figure~\ref{fig:fig2} we present the Tycho's NE Balmer filaments that we observed and spatially resolved with \ghafas, and show spatial variation of the median NL width (\wnl\ $\in [35,\,72]$\,\kms) and IL flux fraction (\fil\ $\in [0,\,0.42]$), computed as described in ii). The distribution of all medians and their 95\% confidence intervals are given in the top row in Figure~\ref{fig:fig3} as solid and dotted histograms, respectively. In 24\% of the bins we find a significant evidence for IL, while 18\% of the bins have significant Bayesian evidence for a split in the NL. As expected, posteriors of the global (cross-bin median) parameters are significantly narrower than the constraints provided from an individual bin alone (bottom row in Figure~\ref{fig:fig3}). We found the median and the 95\% confidence intervals of the cross-bin median posteriors on the following parameters: 
\wnl\ $= (54.8 \pm 1.8)$\,\kms, \fil/\fnl\ $= (0.41 \pm 0.07)$, \wil\ $= (180.5 \pm 14.3)$\,\kms, and \dnl\ $= (38.5 \pm 5.1)$\,\kms.

\section{Conclusions and Summary}

We presented observations of the Tycho's Balmer filaments in the NE rim, where for the first time, we spatially resolved the filaments, covered them in their entirety, and spectroscopically resolved the narrow \Ha\ component. We also improved the analysis by applying Bayesian inference to obtain reliable parameter estimates and uncertainties, and to quantify the evidence for models with multiple line components.

Our results show that the \wnl\ is much larger than 20\,\kms\ in the entire NE rim even when the NL is split and modeled accordingly. We find that 18\% of the bins show significant evidence for double-NL models. These findings can currently only be explained by efficient CR acceleration in the NE rim confirming previous results in the Tycho's 'knot\,g' by \cite{ghava00} and \cite{lee07}. The heating in the precursor broadens the NL widths to $\approx$\,55\,\kms, and the momentum transfer introduces two NLs in the observed projected shock emission separated by $\approx$\,38\,\kms\ on average. As the amount of neutrals in the ambient medium around the remnant changes, we expect the NL width to vary across the filaments as we see in Figure~\ref{fig:fig2}. This effect occurs because the pre-shock gas is heated to higher temperature (larger \wnl) if the medium has more neutrals as the ion-neutral damping of magnetic waves excited by CRs is more efficient.

Finally, significant evidence for an IL presence is measured in 24\% of the bins, with \wil\ $\approx$\,180\,\kms\ and \fil/\fnl $\approx$\,0.41. Detection of the IL indicates presence of a BN precursor. Therefore, our observations and analysis reveal the interplay of shock precursors in the NE rim of Tycho.

\end{document}